\documentclass[reprint,amsmath,amssymb,superscriptaddress]{revtex4-1}

\usepackage{graphicx}
\usepackage{dcolumn}
\usepackage{bm}
\usepackage{color}
\usepackage{ulem}

\begin{document}


\title{First-principles study of the anisotropic magneto-Peltier effect}


\author{Keisuke Masuda}
\affiliation{Research Center for Magnetic and Spintronic Materials, National Institute for Materials Science (NIMS), Tsukuba 305-0047, Japan}
\author{Ken-ichi Uchida}
\affiliation{Research Center for Magnetic and Spintronic Materials, National Institute for Materials Science (NIMS), Tsukuba 305-0047, Japan}
\affiliation{Center for Materials Research by Information Integration, National Institute for Materials Science (NIMS), Tsukuba 305-0047, Japan}
\affiliation{Department of Mechanical Engineering, The University of Tokyo, Tokyo 113-8656, Japan}
\affiliation{Center for Spintronics Research Network, Tohoku University, Sendai 980-8577, Japan}

\author{Ryo Iguchi}
\affiliation{Research Center for Magnetic and Spintronic Materials, National Institute for Materials Science (NIMS), Tsukuba 305-0047, Japan}
\author{Yoshio Miura}
\affiliation{Research Center for Magnetic and Spintronic Materials, National Institute for Materials Science (NIMS), Tsukuba 305-0047, Japan}
\affiliation{Center for Materials Research by Information Integration, National Institute for Materials Science (NIMS), Tsukuba 305-0047, Japan}
\affiliation{Center for Spintronics Research Network, Osaka University, Toyonaka, Osaka 560-8531, Japan}


\date{\today}

\begin{abstract}
We study theoretically the anisotropic magneto-Peltier effect, which was recently demonstrated experimentally. A first-principles-based Boltzmann transport approach including the spin-orbit interaction shows that Ni has a larger anisotropy of the Peltier coefficient ($\Delta \Pi$) than Fe, consistent with experiments. It is clarified that spin-flip electron transitions due to the spin-orbit interaction are the key in the mechanism of the large anisotropic magneto-Peltier effect. Using our method, we further predict several ferromagnetic metals with much larger $\Delta \Pi$ than that of Ni.
\end{abstract}


\maketitle

\section{\label{introduction} introduction}
The spin-orbit interaction (SOI) plays a key role in mechanisms of various spintronic phenomena, such as the spin Hall effect \cite{1971Dyakonov-PL}, the Rashba--Edelstein effect \cite{1984Bychkov-JETPLett,1990Edelstein-SSC}, magnetic anisotropies \cite{2017Dieny-RMP}, and the anisotropic magnetoresistance (AMR) \cite{1946Bozorth-PR,1954Karplus-PR,1975McGuire-IEEE}. Among them, a transport phenomenon strongly coupled with the magnetization is the AMR in ferromagnets, where the electrical resistivity depends on the relative angle between the charge current and the magnetization owing to the SOI acting on spin-polarized charge carriers.

\begin{figure}
\includegraphics[width=6.0cm]{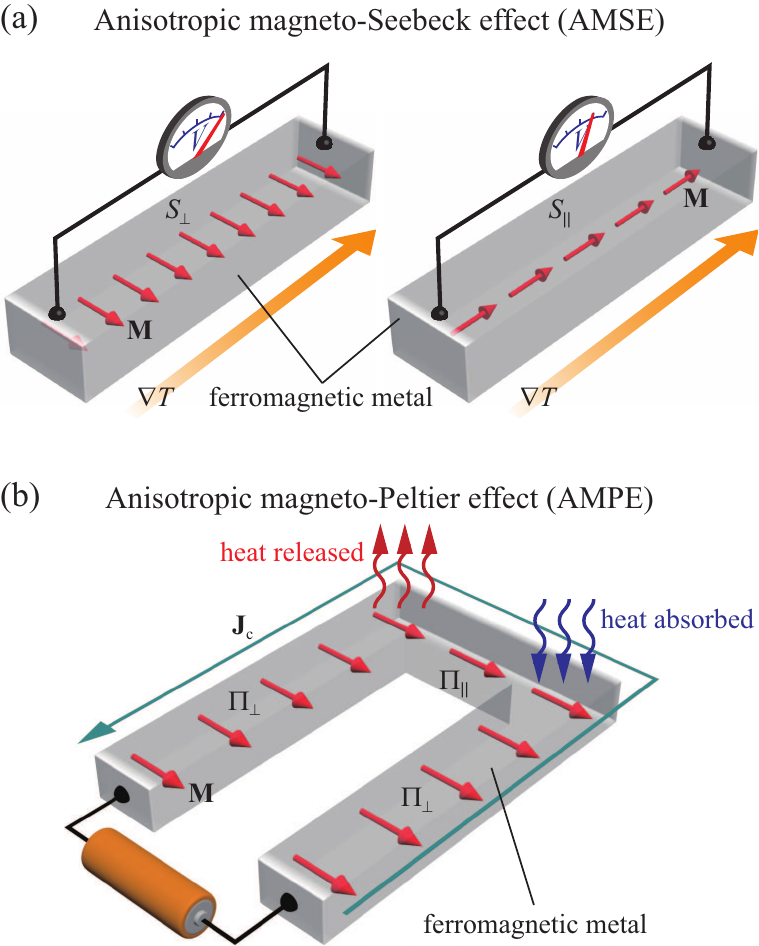}
\caption{\label{Fig1} Schematic illustrations of (a) the AMSE and (b) the AMPE.}
\end{figure}

Similar to the AMR, thermoelectric coefficients also depend on the direction of the magnetization. The Seebeck coefficient $S$ in a ferromagnet is dependent on the relative angle between the directions of the thermal gradient $\nabla T$ and the magnetization ${\bf M}$ [see Fig. \ref{Fig1}(a)], which is called the anisotropic magneto-Seebeck effect (AMSE) \cite{2014Boona-EES,1957Jan-SSP,1976Grannemann-PRB,2006Wegrowe-PRB,2011Avery-PRB,2012Mitdank-JAP,2012Avery-PRL,2012Avery-PRB,2013Schmid-PRL,2017Reimer-SciRep,1954Miyata-JPSJ,2016Watzman-PRB}. The reciprocal effect of the AMSE called the anisotropic magneto-Peltier effect (AMPE), in which the Peltier coefficient $\Pi$ depends on the relative angle between the charge current ${\bf J}_{\rm c}$ and ${\bf M}$, has also been investigated recently \cite{2016Das-PRB,2018Uchida-Nature,2018Das-arXiv}. Uchida {\it et al}. \cite{2018Uchida-Nature} directly observed temperature change due to the difference in the Peltier coefficient $\Delta\Pi=\Pi_{\parallel}-\Pi_{\perp}$, where $\Pi_{\parallel}$ ($\Pi_{\perp}$) is the Peltier coefficient for ${\bf M}\parallel{\bf J}_{\rm c}$ (${\bf M}\!\perp\!{\bf J}_{\rm c}$) in ferromagnetic metal slabs [see Fig. \ref{Fig1}(b)].

Uchida {\it et al}. \cite{2018Uchida-Nature} found that Ni, Ni$_{95}$Pt$_{5}$, and Ni$_{95}$Pd$_{5}$ exhibit clear AMPE. On the other hand, in their experiments, Fe did not show clear AMPE and Ni$_{45}$Fe$_{55}$ showed only small signals. They also observed the similar tendency in the AMSE measurements; the AMSE signal of Ni was clear but that of Fe was negligibly small \cite{2018Uchida-Nature}. Although both the AMPE (AMSE) and AMR are associated with the SOI, we can find a clear difference in the material dependence between these phenomena; while Fe exhibits the small but finite AMR \cite{1975McGuire-IEEE}, no clear AMPE signal was obtained in Fe in contrast with the case of Ni with a large AMPE coefficient \cite{2018Uchida-Nature}. The origin of such a strong material dependence of the AMPE and AMSE should be clarified; however, no theoretical study has addressed the material dependence of these phenomena so far.

In this study, we investigate theoretically the intrinsic mechanism and the material dependence of the AMPE by analyzing the anisotropy of the Peltier coefficient $\Delta \Pi$ on the basis of the first-principles-based Boltzmann transport approach including the SOI. We show that $\Delta \Pi$ in Ni is much larger than that in Fe, in agreement with experimental observations, and that such a difference in $\Delta \Pi$ comes from the presence of the spin-flip electron transition around the Fermi level in Ni. Using this calculation method, we reveal that $|\Delta \Pi|$ of several ferromagnetic metal alloys containing Pt \cite{1988Leroux-JPhysF} is much larger than that of Ni. Although we focus only on the AMPE in this study, the results can be applied to the AMSE simply by dividing $\Delta \Pi$ by the temperature $T$ on the basis of the Onsager reciprocal relations.

\section{\label{method} calculation method}
In the present analysis, we focus on the intrinsic mechanism of the AMPE by applying the first-principles-based Boltzmann transport approach to bulk ferromagnets. This is because the AMPE (AMSE) discussed in this study occurs in bulk ferromagnets and does not require any interfaces, unlike other phenomena; e.g., the spin-dependent Seebeck and Peltier effects in magnetic nanostructures \cite{2010Slachter-NatPhys,2012Flipse-NatNanotech,2012Bauer-NatMat} and the magneto-Seebeck and Peltier effects in magnetic tunnel junctions \cite{2011Walter-NatMat,2015Shan-PRB,2018Kuschel-JPhysD}.

The electronic structure of each system was calculated by using the full-potential linearized augmented plane-wave (FLAPW) method including the SOI implemented in the {\scriptsize WIEN}2{\scriptsize K} program \cite{Blaha_wien2k}. We employed conventional unit cells for bcc Fe and fcc Ni \cite{note_uc-1}, where we set ${\bf M}$ along the [001] direction [see insets of Fig. \ref{AMPE}(a)]. By applying Boltzmann transport theory \cite{BOLTZTRAP} to the obtained electronic structures, we calculated the Peltier coefficient $\Pi_{\alpha}\, (\alpha=\perp,\parallel)$ given by
\begin{equation}
\Pi_{\alpha}=-\frac{1}{e} \frac{\int \sigma_{\alpha}(\epsilon)(\epsilon-\mu)(-\frac{\partial f}{\partial\epsilon})\, d \epsilon}{\int \sigma_{\alpha}(\epsilon)(-\frac{\partial f}{\partial\epsilon})\, d \epsilon},\label{eq1}
\end{equation}
where $\mu$ is the chemical potential, $f(\epsilon)=\{\exp{[(\epsilon-\mu)/k_{\rm B}T]}+1\}^{-1}$ is the Fermi distribution function, and $\sigma_{\alpha}(\epsilon)=\frac{e^2 \tau}{N}\sum_{i,{\bf k}}v_{\alpha}(i,{\bf k})v_{\alpha}(i,{\bf k})\delta(\epsilon-\epsilon_{i,{\bf k}})$ is the energy-dependent conductivity. Here, $\epsilon_{i,{\bf k}}$ is the eigenenergy with the wave vector ${\bf k}$ in the band $i$, $v_{\parallel}(i,{\bf k})$ [$v_{\perp}(i,{\bf k})$] is the group velocity along the direction parallel (perpendicular) to ${\bf M}$, $\tau$ is the relaxation time assumed to be constant, and $N$ is the number of {\bf k} points used in the summation. The temperature $T$ in the Fermi function was fixed to 300\,K in our analysis to compare with experiments performed at room temperature \cite{2018Uchida-Nature}. We estimated the AMPE from the anisotropy of the Peltier coefficient $\Delta \Pi \equiv \Pi_{\parallel}-\Pi_{\perp}$.

We also analyzed the AMR given by the anisotropy of the electrical resistivity. The AMR ratio is defined as $\Delta \rho/\rho_{\rm av}=(\rho_{\parallel}-\rho_{\perp})/(\frac{1}{3}\rho_{\parallel}+\frac{2}{3}\rho_{\perp})$, where $\rho_{\parallel}$ ($\rho_{\perp}$) is the electrical resistivity when the electric current is parallel (perpendicular) to the magnetization. Note here that $\rho_{\alpha}$ ($\alpha=\perp,\parallel$) is the inverse of the electrical conductivity $\sigma_{\alpha}$ and is formulated as follows in Boltzmann transport approach:
\begin{equation}
\rho_{\alpha}=\frac{1}{\sigma_{\alpha}}=\frac{1}{\int \sigma_{\alpha}(\epsilon)(-\frac{\partial f}{\partial\epsilon})\, d \epsilon}.\label{eq2}
\end{equation}
By comparing Eqs. (\ref{eq1}) and (\ref{eq2}), we see that the numerator of Eq. (\ref{eq1}) gives the difference in the material dependence between the AMPE and AMR, which will be discussed in more detail in the next section.

\section{\label{resultsdiscussion} results and discussion}
\begin{figure}
\includegraphics[width=7.0cm]{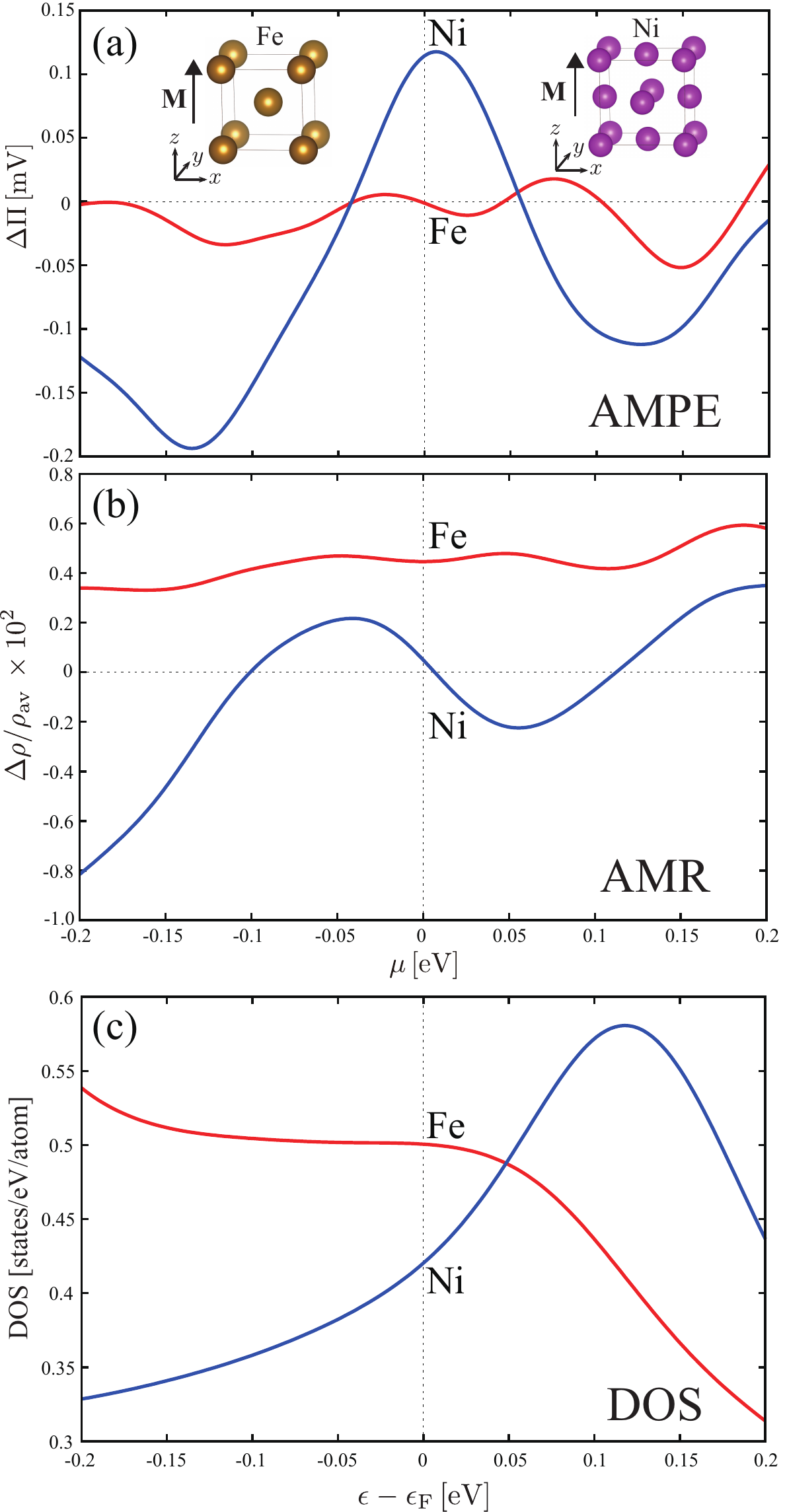}
\caption{\label{AMPE}Calculated $\mu$ dependencies of (a) the anisotropy of the Peltier coefficient $\Delta \Pi=\Pi_{\parallel}-\Pi_{\perp}$ and (b) the AMR ratio $\Delta \rho/\rho_{\rm av}$ for Fe and Ni. (c) The total DOSs of Fe and Ni.}
\end{figure}
\begin{figure}
\includegraphics[width=9.2cm]{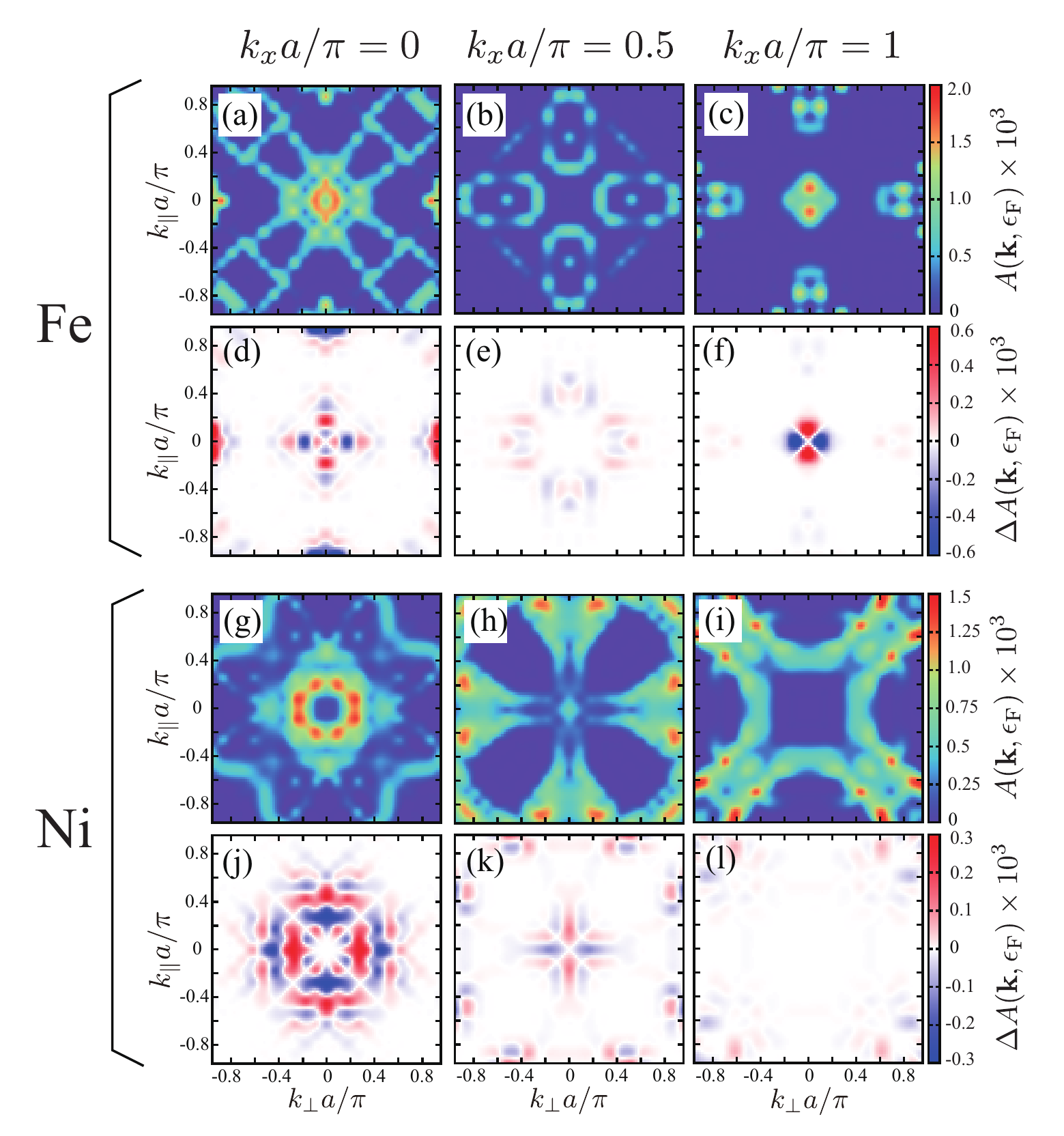}
\caption{\label{spw} (a)--(c) The in-plane wave vector $(k_{\perp},k_{\parallel})$ dependencies of the spectral weight $A({\bf k},\epsilon_{\rm F})$ in Fe at (a) $k_x a/\pi=0$, (b) $k_x a/\pi=0.5$, and (c) $k_x a/\pi=1$. (d)--(f) The in-plane wave vector $(k_{\perp},k_{\parallel})$ dependencies of the anisotropy of the spectral weight $\Delta A({\bf k},\epsilon_{\rm F})$ in Fe at (d) $k_x a/\pi=0$, (e) $k_x a/\pi=0.5$, and (f) $k_x a/\pi=1$. (g)--(i) The same as (a)--(c) for Ni. (j)--(l) The same as (d)--(f) for Ni.}
\end{figure}
\begin{figure}
\includegraphics[width=6.25cm]{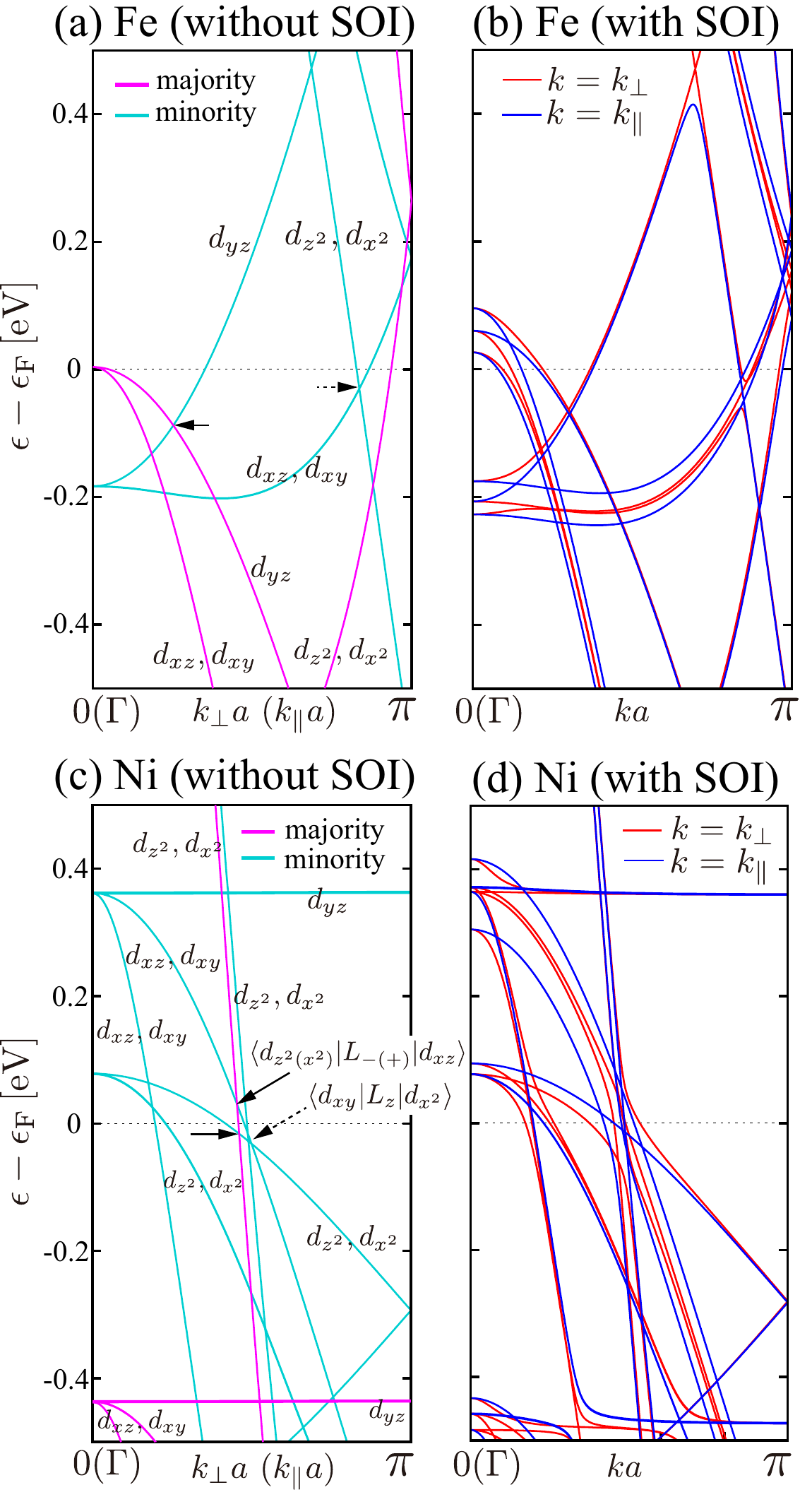}
\caption{\label{band} Band structures along the $k_{\perp}$ and $k_{\parallel}$ lines through $\Gamma$ for Fe in the (a) absence and (b) presence of the SOI. (c), (d) The same as (a) and (b) but for Ni. In panels (a) and (c), $d$ orbitals contributing to each band are indicated on the curve, where $d_{3z^2-r^2}$ and $d_{x^2-y^2}$ are, respectively, abbreviated as $d_{z^2}$ and $d_{x^2}$ for simplicity.}
\end{figure}
\begin{figure}
\includegraphics[width=7.0cm]{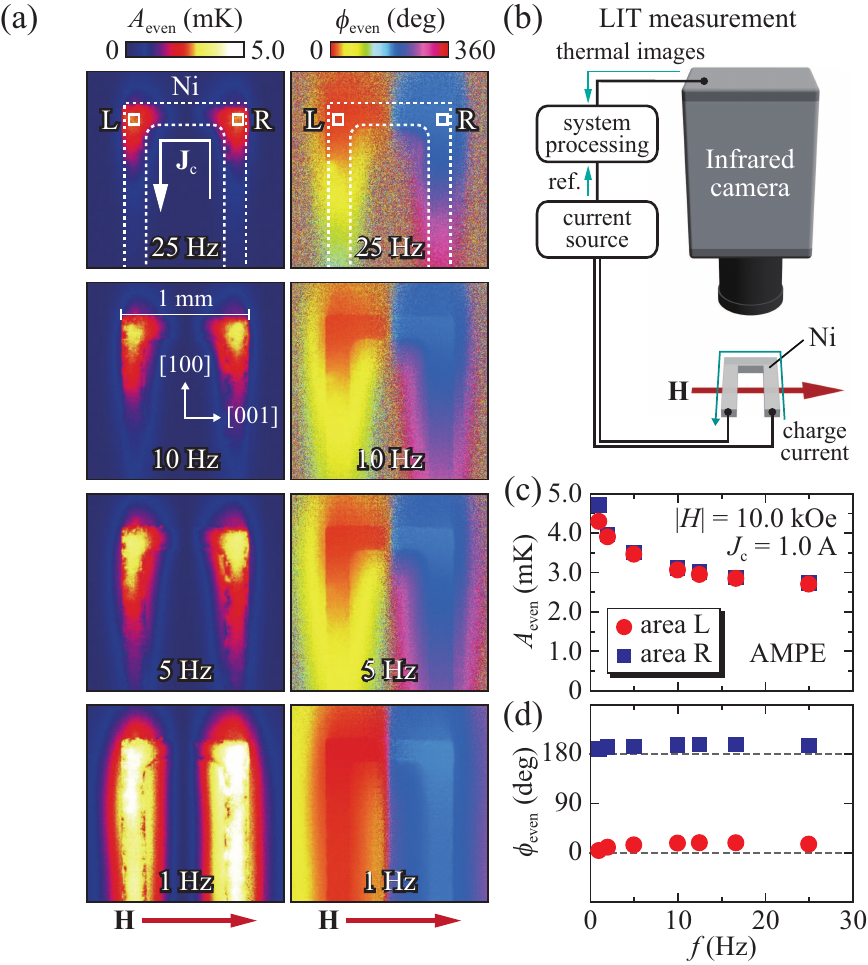}
\caption{\label{Fig_experiment} Experimental results on the AMPE and AMSE in the single-crystalline Ni. (a) Lock-in amplitude $A_{\rm even}$ and phase $\phi_{\rm even}$ images (b) Schematic of the setup in the AMPE experiments. (c),(d) LIT frequency $f$ dependencies of $A_{\rm even}$ and $\phi_{\rm even}$ at the corners L and R of the U-shaped Ni sample.}
\end{figure}
In Fig. \ref{AMPE}(a), we show the calculated anisotropy of the Peltier coefficient $\Delta \Pi$ as a function of the chemical potential $\mu$ for Fe and Ni. Here, $\mu=0$ corresponds to the Fermi level $\epsilon_{\rm F}$. At $\mu=0$, Ni has a $\Delta \Pi$ of 0.114\,mV, which is much larger than that of Fe \cite{note_AMSE-Fe}. These values of $\Delta \Pi$ are consistent with the experimental ones \cite{2018Uchida-Nature}, not only qualitatively but also quantitatively. We also calculated the $\mu$ dependencies of AMR ratios for Fe and Ni, as shown in Fig. \ref{AMPE}(b), where the AMR ratio of Fe is larger than that of Ni around $\mu=0$. Such opposite material dependences of the AMPE and AMR clearly indicate that these phenomena follow different physical pictures, consistent with the findings of previous theoretical studies \cite{2018Popescu-JPhysD,2014Wimmer-PRB}.

To explain the fundamental physical properties of the Peltier coefficient, one can utilize the following approximated expression of Eq. (\ref{eq1}), called the Mott formula \cite{Ashcroft-Mermin}:
\begin{equation}
\Pi \approx -\frac{\pi^2 (k_{\rm B}T)^2}{3e} \frac{1}{D(\epsilon_{\rm F})} \left[ \frac{dD(\epsilon)}{d\epsilon} \right]_{\epsilon=\epsilon_{\rm F}},\label{Mott}
\end{equation}
where $\Pi$ is the Peltier coefficient and $D(\epsilon)$ is the density of states (DOS). This expression suggests that a smaller $D(\epsilon_{\rm F})$ and a larger $[dD(\epsilon)/d\epsilon]_{\epsilon=\epsilon_{\rm F}}$ are better for obtaining a larger $\Pi$ and its anisotropy. As shown in Fig. \ref{AMPE}(c), the DOS of Ni satisfies such conditions.

Our results for the AMR ratios can be understood from the relation that the electrical conductivity is approximately proportional to the DOS at the Fermi level $D(\epsilon_{\rm F})$, which is derived in the same manner as Eq. (\ref{Mott}) \cite{Ashcroft-Mermin}. This relation holds for the present case because we confirmed that the calculated $\sigma_{\perp}$ ($\sigma_{\parallel}$) of Fe is larger than that of Ni, as expected from the DOSs shown in Fig. \ref{AMPE}(c). Owing to such a material dependence in the conductivity, the anisotropy $\sigma_{\perp}-\sigma_{\parallel}$ is also larger in Fe than in Ni, leading to the AMR ratios shown in Fig. \ref{AMPE}(b). Note here that our calculations for the AMR ratios took into account only the intrinsic contribution from the band structures of Fe and Ni. On the other hand, in actual experiments, $s$-$d$ scattering due to impurities provides non-negligible contributions to the conductivity \cite{2012Kokado-JPSJ}, leading to the experimental behavior that the AMR ratio of Ni is larger than that of Fe \cite{1975McGuire-IEEE}. However, in the case of the AMPE, not only the DOS itself but also its derivative give a large contribution to the Peltier coefficient. Therefore, in ferromagnets having a large $[dD(\epsilon)/d\epsilon]_{\epsilon=\epsilon_{\rm F}}$ and a small $D(\epsilon_{\rm F})$, the effect of the $s$-$d$ scattering is relatively weakened. This is a possible reason why our calculated value of $\Delta \Pi$ in Ni agrees quantitatively with the experimental results, even though the effect of the $s$-$d$ scattering is disregarded and the relaxation time is assumed to be constant in the present study.

To obtain further insight into the large $\Delta \Pi$ in Ni, we focus on spectral weights $A({\bf k},\epsilon)=\sum_{i} \delta(\epsilon-\epsilon_{i,{\bf k}})$ \cite{sp-weight} of Fe and Ni. Figures \ref{spw}(a)--\ref{spw}(c) show the two-dimensional wave vector $(k_{\perp},k_{\parallel})$ dependences of $A({\bf k},\epsilon_{\rm F})$ in Fe at $k_x a/\pi=0$, 0.5, and 1, respectively. Here, $k_{\parallel (\perp)}$ represents the wave-vector coordinate parallel (perpendicular) to ${\bf M}$; we set $k_{\parallel}=k_z$ and $k_{\perp}=k_y$ in the present case. Owing to the presence of the SOI, fourfold symmetry of $A({\bf k},\epsilon)$ in the $(k_{\perp},k_{\parallel})$ plane is slightly broken for all $k_x$. This gives the anisotropy of $A({\bf k},\epsilon)$ between $k_{\perp}$ and $k_{\parallel}$ directions, $\Delta A({\bf k},\epsilon_{\rm F}) \equiv A(k_x,k_y,k_z,\epsilon_{\rm F})-A(k_x,k_z,k_y,\epsilon_{\rm F})$, as shown in Figs. \ref{spw}(d)--\ref{spw}(f). From Figs. \ref{spw}(d) and \ref{spw}(f), we see that non-negligible values of $\Delta A({\bf k},\epsilon_{\rm F})$ occur around the $k_{\perp}$ and $k_{\parallel}$ lines through (0,0), which can yield a finite value of $\Delta \Pi$. Figures \ref{spw}(j)--\ref{spw}(l) show the anisotropy $\Delta A({\bf k},\epsilon_{\rm F})$ in Ni calculated from $A({\bf k},\epsilon_{\rm F})$ shown in Figs. \ref{spw}(g)--\ref{spw}(i). We find that Ni has large values of $\Delta A({\bf k},\epsilon_{\rm F})$ especially at $k_x a/\pi=0$ [Fig. \ref{spw}(j)], which distribute broadly around the $k_{\perp}$ and $k_{\parallel}$ lines through (0,0). Such large and distributed $\Delta A({\bf k},\epsilon_{\rm F})$ can be the origin of the large $\Delta \Pi$ in Ni.

To clarify the reason for the difference in $\Delta A({\bf k},\epsilon_{\rm F})$ between Fe and Ni, we next analyzed the band structures along the $k_{\perp}$ and $k_{\parallel}$ lines through $\Gamma$=(0,0,0), as shown in Figs. \ref{band}(a)--\ref{band}(d). Note that these band structures are calculated for conventional unit cells introduced in Sec. \ref{method}. This is why the band structures in Figs. \ref{band}(a)--\ref{band}(d) are seemingly different from the well-known ones calculated for the primitive unit cells \cite{1977Callaway-PRB,1977Wang-PRB}. The band structures for the conventional cells are identical to those obtained by folding the band structures for the primitive cells. If the SOI is absent, we can identify both majority- and minority-spin bands, which have identical dispersions in the $k_{\perp}$ and $k_{\parallel}$ lines [Figs. \ref{band}(a) and \ref{band}(c)]. When the SOI is taken into account, the majority- and minority-spin bands are mixed, and the $k_{\perp}$ and $k_{\parallel}$ lines have different band structures [Figs. \ref{band}(b) and \ref{band}(d)].

The SOI $\xi\, {\bm L}\cdot{\bm S}=\xi\, \left[ \frac{1}{2}(L_{+}S_{-}+L_{-}S_{+})+L_{z}S_{z} \right]$ has two effects on band structures, where ${\bm L}$ and ${\bm S}$ are, respectively, the orbital and spin angular-momentum operators. First, the term $\frac{\xi}{2}\,(L_{+}S_{-}+L_{-}S_{+})$ gives the spin-flip electron transition between majority- and minority-spin bands, leading to the band splitting at their crossing point. In this case, the magnetic quantum numbers $m$ in these bands need to differ with each other by 1 \cite{note_transition_Lpm}. Secondly, the term $\xi\, L_{z}S_{z}$ gives the spin-conserving electron transition between bands with the same spin and the same $m$, also leading to the band splitting \cite{note_transition_Lz}. Since the SOI $\xi\, {\bm L}\cdot{\bm S}$ mainly affects the band structure along the $k_{\parallel}$ line, this interaction gives anisotropic band structures in between $k_{\parallel}$ and $k_{\perp}$ lines, leading to finite $\Delta A({\bf k},\epsilon_{\rm F})$.

Let us discuss each band structure of Fe and Ni in more detail. In the band structure of Fe in the absence of the SOI [Fig. \ref{band}(a)], we have a crossing point between minority-spin bands close to $\epsilon_{\rm F}$ [see the dashed arrow in Fig. \ref{band}(a)]. However, the band splitting due to the SOI is rather weak at this point [Fig. \ref{band}(b)]. We can also find other crossing points between the majority- and minority-spin bands, but they are not close to $\epsilon_{\rm F}$; the closest crossing point from $\epsilon_{\rm F}$ is at $\epsilon-\epsilon_{\rm F} \approx -0.09\,{\rm eV}$ [see the solid arrow in Fig. \ref{band}(a)]. Moreover, at this crossing point, since both majority- and minority-spin bands originate from the same $d_{yz}$ state ($m=\pm1$), the spin-flip electron transition, i.e., the band splitting does not occur when the SOI is taken into account [Fig. \ref{band}(b)]. This is why Fe has small $\Delta A({\bf k},\epsilon_{\rm F})$. On the other hand, Ni has a favorable band structure for large $\Delta A({\bf k},\epsilon_{\rm F})$. First, near the center of the $k$ line, three minority-spin bands cross nearly at a single point around $\epsilon_{\rm F}$ [see the dashed arrow in Fig. \ref{band}(c)]. At this point, since one band includes the $d_{xy}$ component and the other two bands have the $d_{x^2-y^2}$ component, the spin-conserving transition occurs through $L_{z}$, leading to the band splittings. Moreover, we can also find two crossing points between majority- and minority-spin bands at $\epsilon-\epsilon_{\rm F} \approx 0.02\,{\rm eV}$ and $\approx -0.01\,{\rm eV}$, sufficiently close to $\epsilon_{\rm F}$ [see the solid arrows in Fig. \ref{band}(c)]. In addition, at one of them with $\epsilon-\epsilon_{\rm F} \approx 0.02\,{\rm eV}$, the majority-spin band comes from the $d_{3z^2-r^2}$ ($m=0$) and $d_{x^2-y^2}$ ($m=\pm 2$) components and the minority-spin band includes the $d_{xz}$ ($m=\pm 1$) component. Thus, at this point, the spin-flip transition occurs between the majority-spin $d_{3z^2-r^2(x^2-y^2)}$ state and the minority-spin $d_{xz}$ state through the operator $L_{-(+)}$, leading to significant band splittings. We can conclude that, in the case of Ni, both the spin-conserving and spin-flip transitions occur around $\epsilon_{\rm F}$, which is in sharp contrast with the case of Fe with only a weak spin-conserving transition around $\epsilon_{\rm F}$. These transitions give rise to a large anisotropy of the band structure between the $k_{\perp}$ and $k_{\parallel}$ lines [Fig. \ref{band}(d)], which is the origin of the large $\Delta A({\bf k},\epsilon_{\rm F})$ in Ni shown in Fig. \ref{spw}(j). It is well known that the exchange splitting of Ni is smaller than that of Fe \cite{1980Eastman-PRL}. This is the reason why the spin-flip transition is more effective in Ni than in Fe.

\begin{figure}
\includegraphics[width=7.5cm]{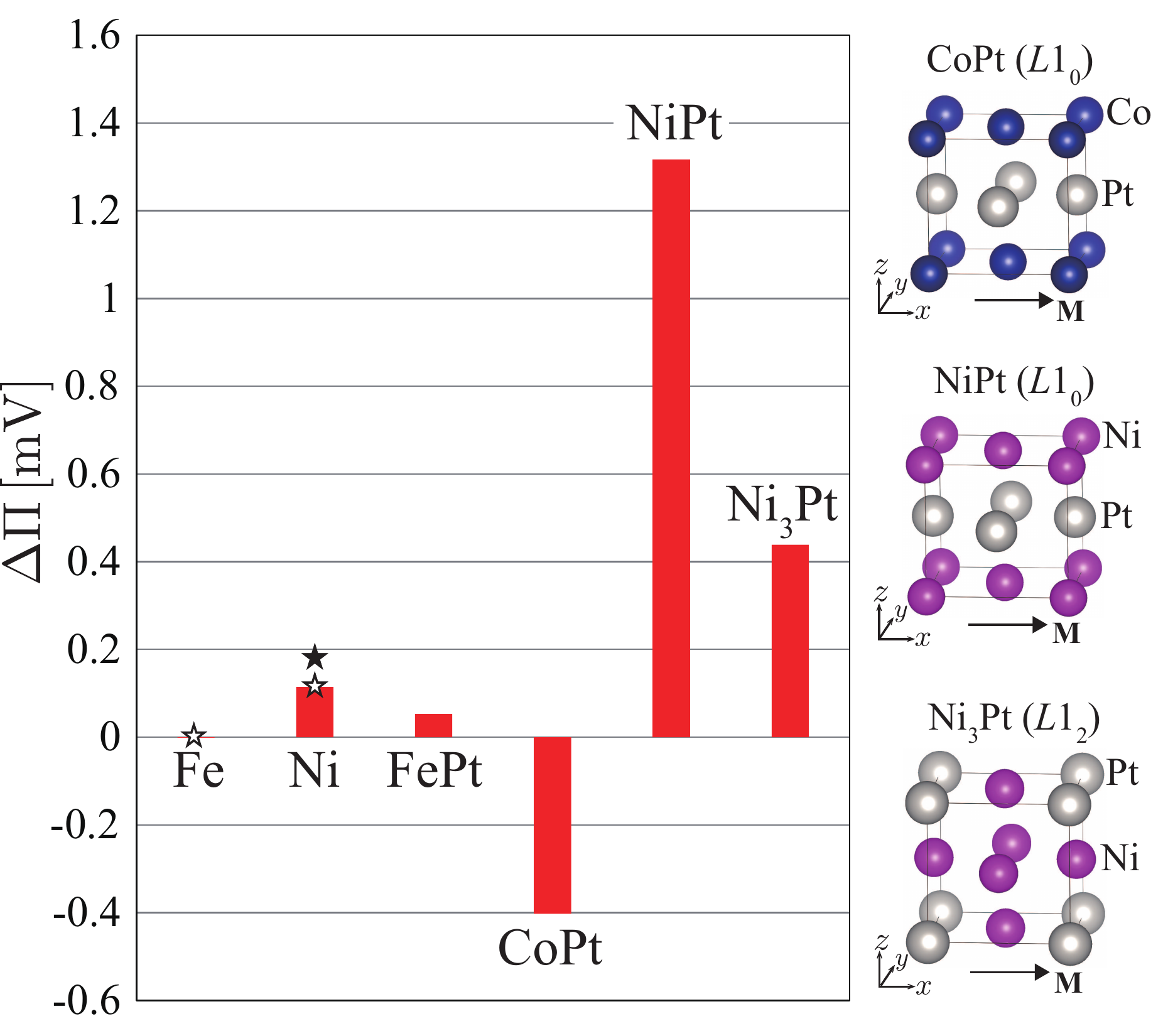}
\caption{\label{AMPE_various-systems} Calculated values of $\Delta \Pi$ at $\mu=0$ for Fe, Ni, FePt, CoPt, NiPt, and Ni$_3$Pt. White and black stars indicate experimental values estimated by the AMPE and AMSE measurements, respectively. Note that Fe does not exhibit clear AMPE in both the calculations and experiments. Crystal structures of CoPt, NiPt, and Ni$_3$Pt are also shown.}
\end{figure}

As mentioned above, our calculated value of $\Delta \Pi$ in Ni agrees with the experimental results in Ref. \onlinecite{2018Uchida-Nature}. However, we assumed single-crystalline Ni in our calculations, although polycrystalline samples were used in the previous study. Thus, we carried out the AMPE experiments using a single-crystalline Ni for direct comparison of our theory with experiments. Using the lock-in thermography (LIT) \cite{2009Straube-APL,2010Breitenstein-Springer,2016Wid-SciRep,2016Daimon-NatCommun,2017Wid-JPhysD,2017Uchida-PRB,2017Daimon-PRB,2017Hirayama-APL}, we observed the distribution of the temperature modulation induced by the AMPE on the surface of a U-shaped single-crystalline Ni slab with ${\bf M}$ along the [001] direction. During the LIT measurements, we applied a magnetic field ${\bf H}$ with the magnitude $H=\pm 10.0\,{\rm kOe}$ and a rectangularly-modulated ac charge current with the amplitude of $J_{\rm c}=1.0\,{\rm A}$ and the frequency of $f=25.0\,{\rm Hz}$, and zero dc offset to the slab [Fig. \ref{Fig_experiment}(b)]. Since the AMPE exhibits an even dependence on the {\bf M} direction \cite{2018Uchida-Nature}, we extract the $H$-even component from the raw LIT images [Fig. \ref{Fig_experiment}(a)], where the LIT amplitude and phase of the $H$-even component are denoted by $A_{\rm even}$ and $\phi_{\rm even}$, respectively. As seen in the top two panels of Fig. \ref{Fig_experiment}(a), the signal is generated around the corners of the U-shaped structure and the sign of the temperature modulation at the corner L is opposite to that at the corner R, which is the behavior expected for the AMPE. In Figs. \ref{Fig_experiment}(c) and \ref{Fig_experiment}(d), we show the $f$ dependences of $A_{\rm even}$ and $\phi_{\rm even}$. By combining these results with numerical calculations based on finite element method \cite{2018Uchida-Nature}, we obtained $\Delta \Pi=0.11\,{\rm mV}$ for single-crystalline Ni, which is almost the same as that for polycrystalline Ni. In Fig. \ref{AMPE_various-systems}, we compare the $\Delta \Pi$ values obtained by our calculations with those by the experiments. Here, white and black stars indicate experimental values estimated from the AMPE and AMSE measurements, respectively. We see that our theoretical values agree well with all the experimental values.

The above calculations remind us of the importance of the SOI in the mechanism of the AMPE. On the basis of this knowledge, we predict promising systems for obtaining large AMPE. We considered $L1_0$-ordered FePt, CoPt, and NiPt, and $L1_2$-ordered Ni$_{3}$Pt \cite{note_ordered-alloy}, since Pt has a strong SOI (Fig. \ref{AMPE_various-systems}). Here, since the [001] direction is special for the $L1_0$ structure, we set ${\bf M}$ along the [100] direction and estimated $\Delta \Pi=\Pi_{\parallel}-\Pi_{\perp}=\Pi_{[100]}-\Pi_{[010]}$. We found that NiPt has a huge $\Delta \Pi$ of 1.31\,mV, which is more than ten times larger than that of Ni. It was also found that CoPt and Ni$_3$Pt have relatively large $|\Delta \Pi|$, which are about four times larger than that of Ni (note that CoPt has a negative $\Delta \Pi$). On the other hand, FePt has a small $\Delta \Pi$ of about half the value in Ni, although FePt is a well-known ferromagnetic metal with strong SOI. Such a nontrivial material dependence of $\Delta \Pi$ clearly indicates that not only a strong SOI but also a small exchange splitting is required for obtaining a large $\Delta \Pi$. Note that, although the largest $\Delta \Pi$ was obtained, NiPt might have a low Curie temperature ($T_{\rm C}\sim 200\,{\rm K}$) as shown in a previous experimental study \cite{1988Leroux-JPhysF}. Thus, to realize huge AMPE at room temperature, CoPt ($T_{\rm C} \gtrsim 800\,{\rm K}$) or Ni$_3$Pt ($T_{\rm C} \gtrsim 300\,{\rm K}$) would be hopeful.

\section{summary}
We gave a microscopic physical picture on the AMPE by calculating the anisotropy of the Peltier coefficient $\Delta \Pi$ on the basis of the first-principles-based Boltzmann transport approach including the SOI. We showed that Ni has a much larger $\Delta \Pi$ than Fe, consistent with recently reported observations on polycrystalline Fe and Ni. By carrying out additional AMPE experiments using single-crystalline Ni, we confirmed that our calculated $\Delta \Pi$ also agrees with the experimental one estimated in single-crystalline Ni, which can emphasize the consistency between our theory and experiments. Analysis of the band structures clarified that the spin-flip electron transition due to the small exchange splitting is the key for the large $\Delta \Pi$ in Ni. Such an insight is important not only for advancing the understanding of the AMPE and AMSE but also for developing researches for other spin-caloritronic phenomena with interconversion between charge and heat currents due to the SOI. We further calculated $\Delta \Pi$ in some ordered alloys including Pt. It was found that $L1_0$-ordered CoPt and NiPt and $L1_2$-ordered Ni$_3$Pt can have huge $|\Delta \Pi|$, which are about several to ten times larger than that of Ni. Our first-principles analysis clarified the microscopic mechanism of the AMPE and predicted hopeful materials to obtain larger AMPE, which is beneficial for developing nanoscale thermal management technologies using electronic and spintronic devices. Further experiments on the temperature dependence of the AMPE would provide more detailed information on the relationship between $\Delta \Pi$ and band structures, which will be addressed in future works.

\begin{acknowledgments}
The authors are grateful to S. Mitani, M. Tsujikawa, K. Nawa, and S. Daimon for many fruitful discussions and helpful comments. This work was supported by JST-CREST ``Creation of Innovative Core Technologies for Nano-enabled Thermal Management''(JPMJCR17I1), the NEC Corporation, and NIMS MI$^2$I. The crystal structures were visualized by using VESTA \cite{2011Momma-JAC}.
\end{acknowledgments}


\begin{thebibliography}{99}
\bibitem{note_uc-1} The $a$-axis lengths of the cells were set to 2.86 and 3.52 \AA\, for Fe and Ni, respectively.
\bibitem{note_ordered-alloy} The $a$-axis lengths of the conventional unit cells were set to 3.86, 3.80, 3.80, and 3.64\,\AA\, for FePt ($L1_{0}$), CoPt ($L1_{0}$), NiPt ($L1_{0}$), and Ni$_3$Pt ($L1_{2}$), respectively. In the systems with $L1_{0}$ structure, the values of $c/a$ were fixed to 0.981, 0.973, and 0.974 for FePt, CoPt, and NiPt, respectively.
\end{thebibliography}

\end{document}